\documentclass[final,3p,times,authoryear]{elsarticle}

\usepackage{amssymb}
\usepackage{hyperref}
\hypersetup{colorlinks=true,citecolor=blue}
\usepackage{graphicx,amsmath}
\usepackage{ifthen}
\usepackage{multirow}
\usepackage{breakurl}
\def\beq{\begin{equation}}
\def\eeq{\end{equation}}
\def\bea{\begin{eqnarray}}
\def\eea{\end{eqnarray}}

\biboptions{round,comma,sort&compress}

\journal{Global Environmental Change}

\begin{document}
\begin{frontmatter}

\title{Modelling complex systems of heterogeneous agents to better design sustainability transitions policy}

\author[RU,CG]{J.-F. Mercure  \corref{cor1}}
\ead{J.Mercure@science.ru.nl}

\author[ce]{H. Pollitt}

\author[KE]{A. M. Bassi}

\author[CG]{J. E. Vin\~uales}

\author[ou]{N. R. Edwards}

\address[RU]{Department of Environmental Science, Radboud University, PO Box 9010, 6500 GL Nijmegen, The Netherlands}

\address[CG]{Cambridge Centre for Environment, Energy and Natural Resource Governance (C-EENRG), University of Cambridge, 19 Silver Street, Cambridge, CB3 1EP, United Kingdom}


\address[ce]{Cambridge Econometrics Ltd, Covent Garden, Cambridge, CB1 2HT, UK}

\address[KE]{KnowlEdge Srl, 2, via San Giovanni Battista, 21057 Olgiate Olona (VA), Italy}

\address[ou]{Environment, Earth and Ecosystems, The Open University, Milton Keynes, UK}

\cortext[cor1]{Corresponding author: Jean-Fran\c{c}ois Mercure}
\fntext[fn1]{Tel: +31 (0) 24 36 53254}

\begin{abstract}

This article proposes a fundamental methodological shift in the modelling of policy interventions for sustainability transitions in order to account for complexity (e.g. self-reinforcing mechanisms, such as technology lock-ins, arising from multi-agent interactions) and agent heterogeneity (e.g. differences in consumer and investment behaviour arising from income stratification). We first characterise the uncertainty faced by climate policy-makers and its implications for investment decision-makers. We then identify five shortcomings in the equilibrium and optimisation-based approaches most frequently used to inform sustainability policy: (i) their normative, optimisation-based nature, (ii) their unrealistic reliance on the full-rationality of agents, (iii) their inability to account for mutual influences among agents (multi-agent interactions) and capture related self-reinforcing (positive feedback) processes, (iv) their inability to represent multiple solutions and path-dependency, and (v) their inability to properly account for agent heterogeneity. The aim of this article is to introduce an alternative modelling approach based on complexity dynamics and agent heterogeneity, and explore its use in four key areas of sustainability policy, namely (1) technology adoption and diffusion, (2) macroeconomic impacts of low-carbon policies, (3) interactions between the socio-economic system and the natural environment, and (4) the anticipation of policy outcomes. The practical relevance of the proposed methodology is subsequently discussed by reference to four specific applications relating to each of the above areas: the diffusion of transport technology, the impact of low-carbon investment on income and employment, the management of cascading uncertainties, and the cross-sectoral impact of biofuels policies. In conclusion, the article calls for a fundamental methodological shift aligning the modelling of the socio-economic system with that of the climatic system, for a combined and realistic understanding of the impact of sustainability policies. 

\end{abstract}


\end{frontmatter}


\section{Introduction}

\subsection{The chilling effect of uncertainty \label{sect:1}}

The starting-point of this article is the need to tackle the uncertainty facing climate policy-making and the related investment decision-making through a more realistic modelling approach.

National and international public policy-making must confront the unprecedented challenge of effectively managing the complex interaction of economic development, energy systems and environmental change \citep{IPCCAR5WGIII}. The effects of stringent climate policies are subject to uncertainty and disagreement, which hinders policy action, and the lack of policy clarity has, in turn, a chilling effect on the private sector's incentives to shift investment towards sustainable options and opportunities. In contexts of damaging policy indecision, investors are often inclined to wait before committing to new long-lived capital investment decisions \citep{IEARealOptions2007}. Meanwhile, carbon budgets are increasingly consumed \citep{IPCCAR5WGI}, and the likelihood of avoiding dangerous climate change is rapidly decreasing. Consensus on desired outcomes achieved through international agreements (COP15 and COP21) urgently needs to be translated into consensus on actions, to identify effective climate policy, and facilitate its rapid global adoption.

At the roots of policy indecisiveness lie conflicts in our understanding of the complex interactions between technology, society, the macroeconomy, and the environment. Policy-makers often consider that important trade-offs exist between, on the one hand, improving the sustainability of the economy and, on the other hand, adequately supporting economic growth. The ensuing reluctance to act contrasts with the signals from an increasing body of reports produced by think-tanks and international organisations arguing that wealth could be generated by new green technology. Decision-makers thus face dissonant signals, causing them to hold back their action. Indeed, successful public (and related private) initiatives of the scale required to stabilise emissions and adapt to climate change have no precedent, and they are unlikely to develop unless the important uncertainty as to their full implications is properly tackled. To make the analysis more intelligible, we identify four major areas where uncertainty contributes to climate policy indecisiveness: (1) the dynamics of technology adoption and diffusion; (2) macroeconomic impacts of low-carbon policies; (3) interaction between human and environmental systems; and (4) policy implementation and effectiveness.  

Solutions to environmental degradation rest on the diffusion of mostly energy-related innovations, technologies and practices throughout industries and between households. In many cases, low-carbon alternatives already exist. However, whether their adoption can be incentivised in time to avoid dangerous environmental change, and whether this is economically or technically possible, are open questions. Similarly, the extent to which such diffusion could support economic development is not well understood. Moreover, it is also unclear whether climate policies may influence access to food, water and energy, and -- if so -- how. Hence, guidance on how to understand the complex interactions between technology, the macroeconomy, and the environment is much needed. 

This article introduces a methodological approach that could significantly improve our ability to anticipate the effects of climate policies, by integrating behavioural and non-equilibrium complexity science and environmental feedbacks into climate policy analysis, with a framework consistent across relevant disciplines.

\subsection{Shortcomings of equilibrium and optimisation-based analysis\label{sect:1.1}}

Equilibrium and optimisation-based approaches (e.g. Cost-Benefit Analysis or CBA, general equilibrium, cost-optimisation), despite their contribution to date \citep[e.g. ][]{GEA2012,Stern2007}, have five main shortcomings for the analysis of the uncertainty identified in the preceding section.

The first shortcoming concerns the concept underpinning these approaches, namely that of an effective social planner coordinating society to minimise total cost or maximise aggregate utility, and the implicit corollary that there exists a unique stable economic equilibrium to which the economy tends to return after exogenous disturbances \citep[e.g. see][]{Mas-Colell1995}. This method leads to approaches that are generally \emph{normative} (i.e. they seek to identify optimal strategies) rather than \emph{positive/descriptive} (i.e. they do not always seek to describe actual system behaviour with a high degree of realism). While a normative approach may appear attractive for policy purposes, it is fundamentally undermined by the fact that no such central coordination exists, while the assumption of optimality neglects critical aspects of economic reality such as unemployment and market disequilbria that act both as drivers of change and opportunities for economic growth. If the economy is assumed to be permanently in an optimal state, then planning for and incentivising change makes little sense.

A second, closely related shortcoming is that equilibrium theories do not sufficiently allow for \emph{the possibility that agents may not be fully rational}. Indeed, equilibrium theories typically involve finding the (inter-temporal) maximum of the aggregate utility function (or, in cost-optimisation models, the minima of total system costs). Such a maximum (minimum) results from the sum of the utility (cost) functions of the underlying agents, who are assumed to carry out an exhaustive ranking of their preferences over all possible products in all existing markets, and to optimise their goods basket choices given an economic context \citep[see][]{Mas-Colell1995}. Underpinning this understanding is the assumption that the aggregate behaviour of a system of utility-maximisers can always be expressed as that of a single average utility-maximising \emph{representative agent} \citep[see][for a critique]{Kirman1992}. However, if one admits that agents do not carry out an exhaustive ranking of preferences (\emph{bounded rationality}) or that agents may influence one another, the utility of the representative agent becomes too complex or impossible to optimise due to increasing feedbacks and emerging complex dynamics. It then becomes unclear whether optimisation methods can be used at all, and whether imposing strictly constant or decreasing returns, in order for models to converge, does not come at the price of losing touch with reality. In other words, humans are not supercomputers optimising their choices over all goods offered in all markets of the planet \citep{Kirman1992}. Agents usually know a small subset of information on goods they desire, and do not desire goods they know nothing of.

Thirdly, equilibrium theories do not capture the possibility that agents \emph{may influence one another, leading to positive feedbacks and increasing returns}. In this regard, the conventional equilibrium perspective may be termed \emph{reductionist}, in the meaning ascribed to this term in complexity theory \citep{Anderson1972}, i.e. the macro system behaviour is aggregated from micro properties, but without considering to a full extent the interactions (including mutual influence) between agents. Following complexity theory, one may consider interactions as \emph{additional elements} that lead to the emergence of additional \emph{collective phenomena}. In economics, the inclusion of multi-agent interactions (specifically, the possibility that agents influence the behaviour of other agents) actually determines the difference between, on the one hand, models of an economy effectively formed of a single agent (or $N$ isolated individuals) and, on the other hand, models of an economy where additional processes due to crowd effects are allowed to emerge, including technology transitions and economic cycles. Allowing for multi-agent interactions is very important in practice because such interactions are at the roots of \emph{al}l self-reinforcing economic processes (crowd effects), and these are neglected in equilibrium economics. In the theory of complex systems, many properties \emph{emerge solely from} the interactions between agents, not from the behaviour of the agents themselves (regardless of whether these are in homogeneous or heterogeneous contexts). This includes important economic phenomena such as the profile of diffusion of innovations, learning-by-doing, expectations in finance and economic fluctuations, trends and fashions, technology lock-ins, and many more. These phenomena exist, but do not stem from microeconomic behaviour of isolated individual agents. Therefore they cannot be studied with a methodological understanding that ignores interactions between agents. With multi-agent interactions, the representative agent may be understood to gain \emph{additional emergent collective behavioural traits} that the underlying agents do not possess when isolated. Thus while there is a good rationale for desiring a macro theory built on micro-foundations, the latter must include not just agent properties (e.g. preferences and income) but also inter-agent interactions, and this is a great -- yet unavoidable -- challenge. 

Because optimisation approaches (optimal growth, computable general equilibrium (CGE), partial equilibrium cost-optimisation) are sensitive to the curvature of their demand and supply or cost functions, as such, they are unfit to fully account for \emph{increasing returns}, understood as self-reinforcing phenomena. This includes for example a decline in prices resulting from cumulative investment (learning), leading to investments that increase the likelihood of more similar investments. In an optimisation context, such positive feedbacks lead to numerical instability of the model solver (due to multiple solutions). Yet, processes with increasing returns are a very important feature of the real world, particularly as regards climate policy. For example, early investments in solar energy may ultimately lead the technology, through learning, to competitiveness and possible market dominance, while a lack of early investments would have confined the technology to niche applications.

A fourth, related shortcoming is the inability of conventional models to account for \emph{multiple solutions and path-dependence}. Indeed, when increasing returns are introduced, several solutions to the optimisation problem emerge, and it becomes unclear which optimum is the correct one. In the investment example above, two different possible future solutions evolve from different early investment decisions. Real-world technological and economic change is thus \emph{path-dependent}. Technology adoption typically follows S-shaped patterns, which stem partly from social influence and interactions, where adoption of innovations increases the likelihood of further adoption of the same innovations \citep{Rogers2010}. Full path-dependency is a key property missing in many current economic models and Integrated Assessment Models (IAMs), the latter combining global energy-economy-climate change phenomena used to assess environmental policy, whether they are based on small (e.g. the DICE model \citealt{DICE}; FUND model \citealt{FUND}; PAGE model \citealt{PAGE2009}) or large datasets (e.g. PRIMES and GEM-E3 models, \citealt{PRIMES,GEM-E3}; TIAM model, \citealt{ETSAP}; MESSAGE model, \citealt{MESSAGE}; AIM model, \citealt{AIM}; REMIND model, \citealt{REMIND}).

Finally, equilibrium theories do not sufficiently account for \emph{agent heterogeneity}. People and firms are represented by the behaviour of a \emph{single representative agent} with rational expectations. This agent is understood as the aggregate collective behaviour emerging from the actions of the underlying agents, which can have distributed preferences \citep{Mas-Colell1995}. But no clear role is ascribed to differences in the distribution of income or other socio-economic and industry parameters, and, following expected utility theory (EUT), only one type of behavioural response exists, namely a decision based on the expected utility associated with different choices times their respective probabilities (assumed to be known). In the real world, at least two types of deviations from EUT typically arise: \emph{behavioural diversity} \citep[variations around a central value, e.g. discrete choice theory,][]{Domencich1975} and \emph{behavioural biases} (systematic deviation from EUT, e.g. prospect theory, \citealt{Kahneman1979}; see also \citealt{Sorrell2011} for a review and taxonomy). Agent heterogeneity can also be interpreted, in a utility maximisation perspective, as some degree of ambiguity in terms of agent perceptions of optimality. Agent heterogeneity is important in the representation of consumer or investor choices, which is critical in the process of the diffusion of innovations, technologies and practices \citep{Knobloch2016}. As can be inferred from standard innovation diffusion theory \citep{Rogers2010}, behavioural response, its diversity, the diversity of social groups, and unequal distribution of information, are precisely what determines actual rates of adoption of innovations. This is standard knowledge in the field of marketing research \citep[e.g.][on product differentiation]{Smith1956}, where firms seek profits out of matching products to diverse consumer profiles. In addition, utility maximisation under budget constraint is known to be an incorrect representation of attitudes towards risk, gains, losses and uncertainty \citep[as shown in prospect theory,][]{Kahneman1979, Tversky1974}.

The shortcomings of equilibrium models in accounting for multi-agent interactions and their self-reinforcing processes (complexity) or for diversity (agent heterogeneity) leads to the exclusion of very important features of reality from the analysis. Indeed, assuming that a simple unique equilibrium solution exists to the CBA of climate change mitigation comes at a price. It neglects both the path-dependency produced by self-reinforcing phenomena (which are at the roots of technology lock-ins, financial bubbles and crises, technology transitions, etc.) and the diversity of agents (which determines the rates of adoption of innovations). Such analytical omissions are reflected in the types of policies that are advocated on that basis. Specifically, they lead to the expectation that a simple internalisation of environmental costs, in the form of a pricing instrument (e.g. a carbon price or tax) will optimally and effectively incentivise technology adoption and diffusion. And, standard CBA fails to explain why, in the real world, such instruments do not play out and deliver as expected \citep{Grubb2014}. In point of fact, the expected optimal success of such mechanisms is but a reiteration of the initial assumptions of the model, rather than a result of the actual analysis: efficient markets, rationality, etc. 

Compared to equilibrium models of the economy, complex, path-dependent models may appear to be less straightforward to interpret, but should ultimately be easier to relate to reality. Complexity is routinely handled in climatology simulations, where it is well understood that small variations between model runs in their starting values (e.g. pressure, temperature, wind velocity) lead to large differences in model outcomes (rainfall, cloud cover, etc) that increase exponentially with simulated time span. This is due to a high degree of non-linear interaction between variables. This aspect is very well characterised and expressed as probability distributions for climate impacts \citep[e.g. in the][summary for policy-makers]{IPCCAR5WGI} Uncertainty increasing with time of projection will arise in almost any domain where complex interactions between system components exist, not least in economics. A key purpose of this article is to argue that economic analysis could benefit from harmonising its methodology with that of the climate sciences.

\subsection{A paradigm shift\label{sect:1.2}}

Integrating both complexity and behavioural sciences as applied to economics would bring path-dependency and agent heterogeneity to the core of the analysis. Complexity science is the cross-disciplinary field that specifically studies properties that emerge from interactions between system components, initially studied in physical, biological and computer sciences \citep[e.g.][]{Anderson1972, Sigmund1993}. By introducing descriptions of how agents behave and interact, theories and models turn from \emph{normative to positive/descriptive} and their methodology evolves from optimisation to simulation, where the analyst relies on `what if' approaches. 

It is understood that in simulations of complex systems, uncertainty plays an important role partly because the theories do not necessarily predict a propensity to return to equilibrium. This is an aspect well understood and managed in the climate sciences, and no inherent reason prevents us from using the same concepts in economics. Indeed, humans behave in unpredictable ways, partly dictated by diverse contexts. The fact that humans have agency or `free will', is a frequent objection raised by social scientists, to the modelling of behaviour. This argument does not contradict our approach, however, for the following reason. Humans do not behave randomly (unlike physical particles); however, while the actions of individuals cannot be predicted, they almost always lie within known bounds, over which statistics can be developed when considering large groups of people (as for physical particles -- e.g. the multinomial logit in social sciences is conceptually identical to the Boltzmann factor in statistical physics). The result is a theory of collective behaviour. In complex systems (e.g. the climate, the economy), both natural and social systems face the same complexity challenges, and their description can be modelled in the same way. 

In practical terms, the complexity-based methodological approach proposed in this article follows an analytical structure where many scenarios are simulated based on possible policy choices, and acceptable results are retained based on a multidimensional range of outcomes that reach given objectives. For example, such an approach can involve filtering scenarios, in multidimensional human and biogeochemical space (e.g. multiple indicators such as those likely to be selected for the recently adopted Sustainable Development Goals), to determine ranges of policy options that enable society to avoid exceeding planetary boundaries \citep{Rockstrom2009,Steffen2015} while ensuring continued human development globally \citep{OXFAM2012}. Ambiguity or conflicting perceptions of optimal policy-making between diverse policy-makers or model users is avoided by leaving subjective outcome value judgments outside of the scientific framework. Indeed, unlike conventional multi-criteria analysis, where optimisation is carried out using subjective weighting of all factors considered given by the modeller, in the context we propose one obtains feasible points in a multidimensional outcome space, and the policy-maker can target a possible subspace in accordance with her/his political platform.

\subsection{Why complexity and heterogeneity are so important for sustainability transitions\label{sect:1.3}}

A sustainability transition inherently involves socio-technical change, which is a highly non-linear, self-reinforcing process with lock-ins that drive expectations, propelled by choices of and adoption by diverse agents with different perspectives and incomes \citep[e.g.][]{Geels2002, Rogers2010}. For such a system, complexity and behavioural sciences provide a suitable analytical framework. Indeed, technology transitions have an inevitably significant influence on the evolution of the economy through productivity and structural change, which occurs with the development of new firms and industries, and the destruction of others \citep[e.g.][]{Arthur1989,Freeman2001, Perez2001, Schumpeter1934, Schumpeter1939}. At the same time, an economic transformation towards higher or lower sustainability takes place in direct interaction with the environment and its biogeochemical cycles. Simultaneous representations of these four domains (technology, society, the macroeconomy and the environment) cannot reliably be optimised even on the largest super-computers. This is so partly because diverse agents will have conflicting definitions of optimality. 

By contrast, in models capable of accounting for agent diversity and multi-agent interactions, interactions among technology, society, the macroeconomy and the environment can indeed be simulated, much like in models used to simulate the climate. Such models would need to be applied within a framework of uncertainty analysis, in that they would determine the likely outcomes (within uncertainty bounds) of different policies or policy packages as applied to interacting heterogeneous agents. Such knowledge would provide policy-makers with a much more realistic platform to make decisions, and a platform that could be more easily tailored to the specific features of a given policy context (e.g. a region, a country, a sector, etc.).

A complexity paradigm has been suggested as a practical approach to policy problems \citep{Probst2014}. A field of research is also emerging for modelling socio-technical regime transitions (e.g. \citealt{Kohler2009, Holtz2011, Holtz2015, Mercure2015, Safarzynska2010, Safarzynska2012}, see also \citealt{Turnheim2015}), in which `core characteristics of transitions' are itemised as (i) `multi-domain interactions', (ii) `path-dependency', and (iii) `drivers and self-reinforcement of change' \citep{Holtz2011}. System Dynamics \citep{Sterman2000} and agent-based methods \citep{Tesfatsion2002, Buchanan2009}, however, currently have scalability challenges and thus have not yet been integrated into IAM scale analysis, or into other forms of global macro models. It is not necessary -- in principle -- to model every individual agent across sectors and its interactions. Equivalent but simpler statistical models, for instance of technology adoption by interacting agents, can be readily scaled up \citep{Mercure2012, Mercure2014, Mercure2015}, and non-equilibrium models of the global economy already exist \citep[e.g.][]{E3MEmanual, Meyer2007}. At the next level of complexity, representative cohorts or ensembles can be simulated to obtain system statistics, while still simulating orders of magnitude fewer individuals than in the real system. Such approaches are used successfully in global ecosystem models such as HYBRID \citep{Friend2000} and LPJ-GUESS \citep{Smith2001} as well as in intergenerational investment and savings modelling \citep{Miles1999, Rangel2003, Gokhale2002}.

In this paper, we provide some concrete examples of how complexity and behavioural sciences can be used for the assessment of sustainability policies, with emphasis on model uncertainty analysis, in order to build a powerful approach for next-generation public policy analysis. We identify four key areas of climate policy analysis where uncertainty is high and where the proposed methodological approach would be particularly useful: dynamics of technology adoption and diffusion (section~\ref{sect:2.1}); macroeconomic impacts of low-carbon policies (section~\ref{sect:2.2}); interaction between human and environmental systems (section~\ref{sect:2.3}); and policy implementation and effectiveness (\ref{sect:2.4}). We then provide four specific applications of the proposed methodological shift in connection with: the diffusion of transport technology (section~\ref{sect:3.1}); impact of low-carbon investment on income and employment (section~\ref{sect:3.2}); processes with cascading uncertainty (\ref{sect:3.3}); and cross-sectoral impacts of biofuels policy (\ref{sect:3.4}).

\section{Four key areas of climate policy where uncertainty is high\label{sect:2}}

\subsection{Technology adoption and diffusion \label{sect:2.1}}

Both complexity and agent heterogeneity are important to analyse the dynamics of technology adoption and diffusion. The diversity of consumers and investors influences the choice of environmental technology. Consumers' choice of certain technologies, such as vehicles, takes place within contexts of distributed income that span several orders of magnitude \citep[e.g.][]{MercureLam2015}. The diversity of incomes, social groups and attitudes is known to determine the rates and profiles of diffusion of innovations \citep[early and late adopters, etc, see][]{Rogers2010}. For present purposes, that means that agent heterogeneity matters in real life, and it must therefore be integrated into models. Modelling the impacts of policy incentives to the consumer using averages over too few consumer group parameters can indeed be misleading, since differences in income and other socio-economic parameters across consumers lead to highly diverse consumption habits. 

Meanwhile, interactions between agents are also crucial to consider when analysing technology adoption, since technology generally features \emph{increasing returns to adoption}: the adoption of technology by agents is increasingly more probable the more agents adopt and use the technology. Transitions build upon themselves with dynamics following a pattern comparable to that of infectious diseases \citep[see][]{Mansfield1961}. Thus, increasing returns may generate resistance to change in socio-technical regimes (lock-ins), but they may also give momentum to technological transitions (\citealt{Anderson1989}, pp 9-10, \citealt{Arthur2014, Arthur1989}). This aspect is treated in detail in the traditions of evolutionary economics \citep[e.g.][]{Freeman1988} and transitions theory \citep[e.g.][]{Geels2002}.

For obvious reasons, anticipating how diverse agents may respond to different policies has elicited great interest \citep[see][]{Grubb2014}. The impact of incentives to consumers with differing incomes and social groups has been studied in detail from the perspective of marketing \citep[empirically, e.g.][]{Bass1969, Fisher1971, McShane2012}, anthropology \citep{Douglas1979} and behavioural economics (discrete choice modelling, e.g. \citealt{Ben-Akiva1985, Domencich1975}, and behavioural response, e.g. \citealt{Kahneman1979, Tversky1974}). However, the contribution of this research to the understanding of climate policy has so far been mostly overlooked in climate change mitigation modelling or environmental assessment research. Their use thus remains a promising but emerging field \citep[e.g.][]{Kohler2009, Axsen2009, Giraudet2012, Rivers2006}. \cite{Grubb2014} emphasises the need for future research on behavioural aspects of emissions reductions. Insufficient efforts have been devoted to understanding the aggregate behavioural response to sustainability policy instruments, and this is related to a possibly overstated expectation that an efficient equilibrium should emerge on its own in technology markets.

\begin{figure}[t]
	\begin{center}
		\includegraphics[width=.5\columnwidth]{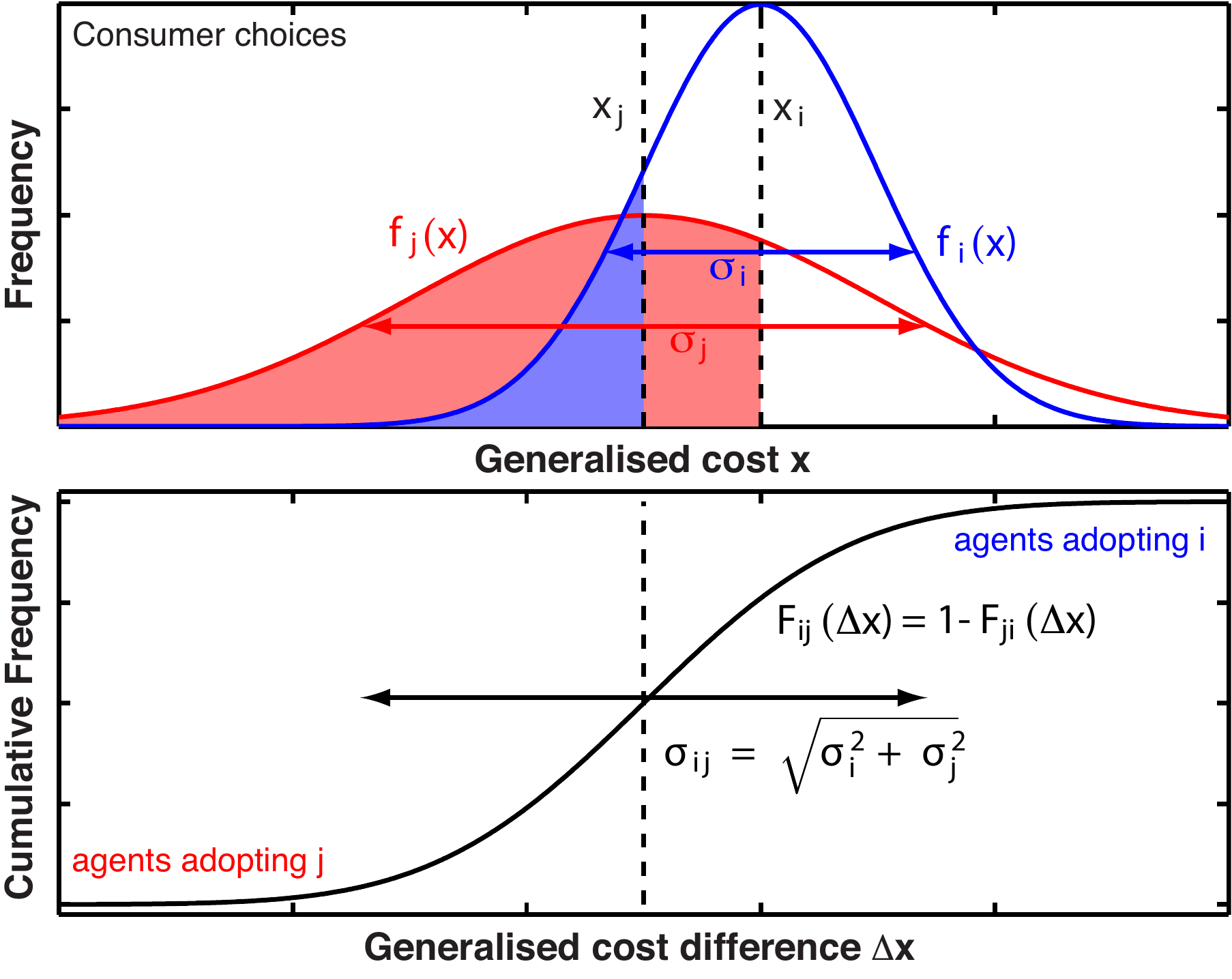}
	\end{center}
	\caption{Schematic representation of a binary logit, reproduced from \citep{Mercure2015}. Agent perceptions of the generalised cost (i.e. including non-pecuniary terms) of choice options vary around central values. In a choice between two options with distributed costs, neither is adopted by all agents; each gets a fraction of agent choices according to a comparison between perceived cost distributions, and the rate of adoption depends strongly on the diversity of agent perceptions.}
	\label{fig:Figure0}
\end{figure} 

Empirical marketing research literature shows that technology substitution in many different contexts follows simple S-shaped diffusion profiles \citep[e.g.][]{Fisher1971, Grubler1999, Mansfield1961, Marchetti1978}. More generally, the competition between several technologies for market space can be described by coupled Lotka-Volterra systems \citep{Bhargava1989, Karmeshu1985}. More recently, it has been shown that Lotka-Volterra systems can be derived from simple statistics of industrial dynamics \citep{Mercure2015}. This can equivalently be expressed with the `replicator dynamics' system of evolutionary theory: \emph{natural selection is carried out by the consumer, who filters successful innovations based on their fitness to the market, while entrepreneurs strive to improve their products in order to increase their fitness by better matching consumer taste} \citep[e.g.][]{Safarzynska2010}. This natural selection, however, involves decision-making by consumers under bounded rationality, who are naturally highly diverse, and the diversity of consumers drives product differentiation and increasing product diversity (for example with private vehicles, see section~\ref{sect:3.1} below).

Equilibrium/optimisation models neglect agent heterogeneity and interactions, as they look at average agents in isolation facing choices (with perfect information), rather than at socio-technical systems where agents influence what other agents do. Agent heterogeneity could, however, be represented to some extent in models by introducing statistical distributions over agent perspectives (see Figures~\ref{fig:Figure0} and~\ref{fig:Figure2}: preferences are often simply distributed), instead of using mean values only. In such a framework, the modeller disposes of \emph{both} mean values as well as ranges. Then, when evaluating aggregate agent propensities towards technology substitution, one simply faces comparisons between probability distributions, as for instance in a binary logit. Chains of binary logits enable to define agents with \emph{bounded rationality} and to model diffusion dynamics, not typically done in current models \citep{Mercure2015}.

This is an area where our methodological shift may advance the understanding of sustainability policy by moving the research focus from the generation of `desirable' energy sector storylines (policy formulation) to the forecasting of `likely' policy outcomes (policy assessment) based on existing technology, knowledge of the market, and technology diffusion dynamics. More fundamentally, the proposed methodology disentangles normative and descriptive analysis as useful but distinct perspectives \citep[see][ for an extensive discussion]{Mercure2014}. However, as is the case with any model involving non-linear dynamics, uncertainty over model outcomes due to uncertainty over parameters increases exponentially with modelling time span, an issue that needs to be addressed (see section~\ref{sect:2.3} below).

\subsection{Green growth: macroeconomics and the finance of innovations\label{sect:2.2}}

Complexity and agent heterogeneity are also important to understand the interactions between technology diffusion, lenders' expectations and macroeconomic fluctuations, which drive economic development/growth and are affected by climate change mitigation policies. Theories based on utility-maximising non-interacting agents assume full employment of economic resources. Under full employment, policies that lead to a re-allocation of resources from an optimal equilibrium starting-point are unavoidably detrimental to economic performance. However, full employment is never observed in reality. In real life, expectations as to the performance and return of innovations drive most investment, and ultimately, economic development and growth. Importantly, expectations as a process arise from multi-agent interactions \citep[e.g. see the artificial stock market by][]{Arthur1997}. 

Policies for the diffusion of low-carbon innovations present significant opportunities for both the private sector and job creation \citep[e.g.][]{Blyth2014, CBI2014}. But the feedback between the diffusion of new sustainability technologies and economic development/growth is poorly understood and reported. There is an apparent contradiction between observations of highly lucrative activity arising in successful low-carbon ventures (e.g. Tesla electric and Toyota hybrid cars, wind turbines, solar photovoltaic), and the perception that the use of lowest-cost fossil energy with conventional technology is indispensable for development/growth. This type of dichotomy pervades climate policy-making.

In conventional equilibrium economics, full employment (which results from maximising the utility of the representative agent) entails that all resources are currently allocated in the economy in the most productive way they can be. This is contradicted by empirical observation \citep[e.g. as reviewed by][]{Grubb2014}, which shows that different countries lie at varying distances from theoretically defined efficiency or productivity frontiers. Effectively, the use of economic resources depends upon the economic development trajectory followed by an economy (investment, fixed and human capital and labour, \emph{ibid}). Moreover, in equilibrium-based theories, savings are understood as a share of fixed national income (GDP), equal to investment, resulting from a choice by agents between spending on consumption in the present or in the future. Thus if a fixed share of income is allocated to investment, naturally, all firms compete for this fixed allowance (crowding-out), equilibrating supply and demand for finance. Since in equilibrium investment and labour is fixed, equilibrium-based theories cannot reproduce financial fluctuations or crises, such as those observed since 2007, or involuntary unemployment, as currently observed in many parts of Europe \citep{Eurostat2015}. 

In a broader complexity/non-equilibrium economics perspective, income is not fixed, and thus investment does not need to be a constant share of the `welfare of the representative agent'. Instead, causality is reversed: income heavily depends upon investment, and investment decisions depend upon decisions by financial institutions. This link, however, opens the door to fluctuations: interacting investors in technology ventures can influence each other into frenzies or panics. Indeed, the finance of technology and innovation leads to productivity change \citep[a phenomenon demonstrated already in the 1950s by][]{Solow1957} but also to speculation and bubbles \citep{Keynes1936}, which, in turn, create economic cycles \citep[see][]{Perez2001, Freeman2001, Schumpeter1934, Schumpeter1939}. Investment fluctuations heavily influence the level of economic development. This dynamic was at the roots of the great recession \citep{Keen2011}, after which quantitative easing was used to offset the reluctance of financial institutions to provide credit. This dynamic cannot be understood without taking into account the interactions between technology diffusion, lender expectations and macroeconomic fluctuations.

More realistic economic modelling that allows for variable amounts of finance and investment is possible if one abandons the restrictive assumptions of equilibrium approaches (imposing a theory of non-interacting agents that are all simultaneously employed or not employed) with their assumed full employment of economic resources. In a theoretical approach that allows for crowd effects to emerge, the amount of investment in the economy is determined by investment decisions by lenders acting upon expectations of return. This is not a new theoretical approach. In point of fact, it is at the roots of both Schumpeterian \citep{Schumpeter1934, Schumpeter1939} and Keynesian \citep{Keynes1936} forms of demand-led economics. A post-Schumpeterian or post-Keynesian perspective also allows for fluctuating amounts of \emph{money} in the economy, and indeed, this is what is observed \citep[called \emph{endogenous money}, as explained by the Bank of England, see][]{McLeay2014}.

As understood in the tradition of Schumpeterian economic history, the emergence of new technologies is characterised by important increasing returns to investment and adoption, and it involves strongly path-dependent cross-sectoral spill-overs that often lead to economy-wide activity accelerators \citep{Freeman2001}. For example, in the industrial revolution, new textile machinery required better iron and steel. Investment in iron and steel led to the emergence of an industry as well as cost reductions that reshaped the whole `\emph{design space}' for other products to be made out of those materials more cheaply (\emph{ibid}). This, in turn, enabled many related and clustered innovations to emerge and economically reach the marketplace \citep[a statistical analysis of the process of clustering of innovations is given by][]{Arthur2006}. In the Schumpeterian economic tradition, the clustering of investment stemming from the clustering of innovation leads to both economic prosperity and depression periods in alternation \citep{Freeman2001, Perez2001, Schumpeter1939}. Risks of financial crises arise when finance is raised using financial assets as collateral \citep{Keen1995, Keen2011}, a phenomenon known to have taken place periodically over history \citep{Perez2001}, including in recent years \citep{Keen2011}. 

Formal modelling of path-dependent, non-equilibrium macroeconomics is possible on the basis of two assumptions. Firstly, under the principle of \emph{cumulative causation of knowledge accumulation}, productivity change is described by sets of aggregate sectoral learning curves where productivity growth is derived from cumulative investment and is roughly consistent with the trends observed in economic history, e.g. in \cite{Kaldor1957, Schumpeter1934}. Secondly, income must be allowed to depend upon investment, instead of the reverse. This understanding is path-dependent and implicitly integrates multi-agent interactions by allowing increasing returns: learning (i.e. knowledge accumulation), investment fluctuations (through expectations), economic cycles and technology diffusion (through social interactions). One example of such a model is E3ME/G \citep[see \burlalt{www.e3me.com}{http://www.e3me.com},][]{E3ME}, which derives a closed set of functional relationships using regressions on economic data, and thus enables the evaluation of employment, instead of remaining constrained by the full employment assumption. Other similar models exist, including GINFORS \citep{Meyer2013}. However, so far, no large-scale model has included detailed dynamics of the financial sector \citep{PollittMercure2015}.

With regard to climate change mitigation, it is well established that an amount of investment in the energy sector larger than in a business-as-usual scenario will be necessary \citep[e.g.][]{IEAWEO2012}. This scale of investment in new technology sectors is likely to result in substantial economic reallocations due to cross-sectoral spill-overs (e.g. new materials, new design and engineering methods, etc). Importantly, in models without the constraint of full-employment of resources, capital and labour, the impact can be either beneficial or detrimental, depending on many factors, such as the trade balance, energy imports, international competition and relative prices. When the implications are beneficial, the transition can be termed as \emph{green growth}, where new investment and employment are generated from technology policy, although possibly with important relative price changes. In section~\ref{sect:3.2}, we discuss an example where climate change mitigation policy leads to increased employment and GDP as a result of enhanced investment in the electricity sector. This said, technology policy does not necessarily lead to beneficial impacts because such policies can take many forms. In an equilibrium-based theory, however, the possibility of beneficial outcomes is entirely ruled out, and by assumption rather than as a result of calculations. Indeed, equilibrium models \emph{always} predict detrimental impacts for climate mitigation efforts \citep[e.g. as in all the models chosen in the 5th assessment report of the IPCC, see][Ch. 6, p.450]{IPCCAR5WGIII}, such that the debate is framed in terms of \emph{burden}, and economic opportunities of climate policies are not extensively examined (\citealt{Grubb2014} see, however, \citealt{Ekins2011, Lee2015}). This is also the case of partial equilibrium models, where, such as in the case of the World Energy Model utilized by the International Energy Agency for its World Energy Outlook report, the investment cost of interventions is estimated, but it is not compared to policy-induced avoided costs nor the impact of these interventions on the macroeconomy. The result is a partial analysis that emphasizes only the costs, and not the benefits of intervention \citep{UNEP2011}. 

Thus, and importantly, we see that the incorporation of multi-agent interactions (by accounting for feedback loops, delays and nonlinearity) leads to key consequences for theory and its application: in a theory where investment is determined by expectations, information flow between variables runs from variable or fluctuating savings/investment to income, back to investment, and allows for beneficial impacts of technology policy while in equilibrium-based theories, information flows from fixed income to investment, and beneficial impacts are ruled out. Given the scale of the socio-economic transformation required to address climate change, a realistic understanding of whether a sustainability transition involving mitigation policies favours or hinders economic development/growth is key for sound policy-making.
 
\begin{figure}[t]
	\begin{center}
		\includegraphics[width=.7\columnwidth]{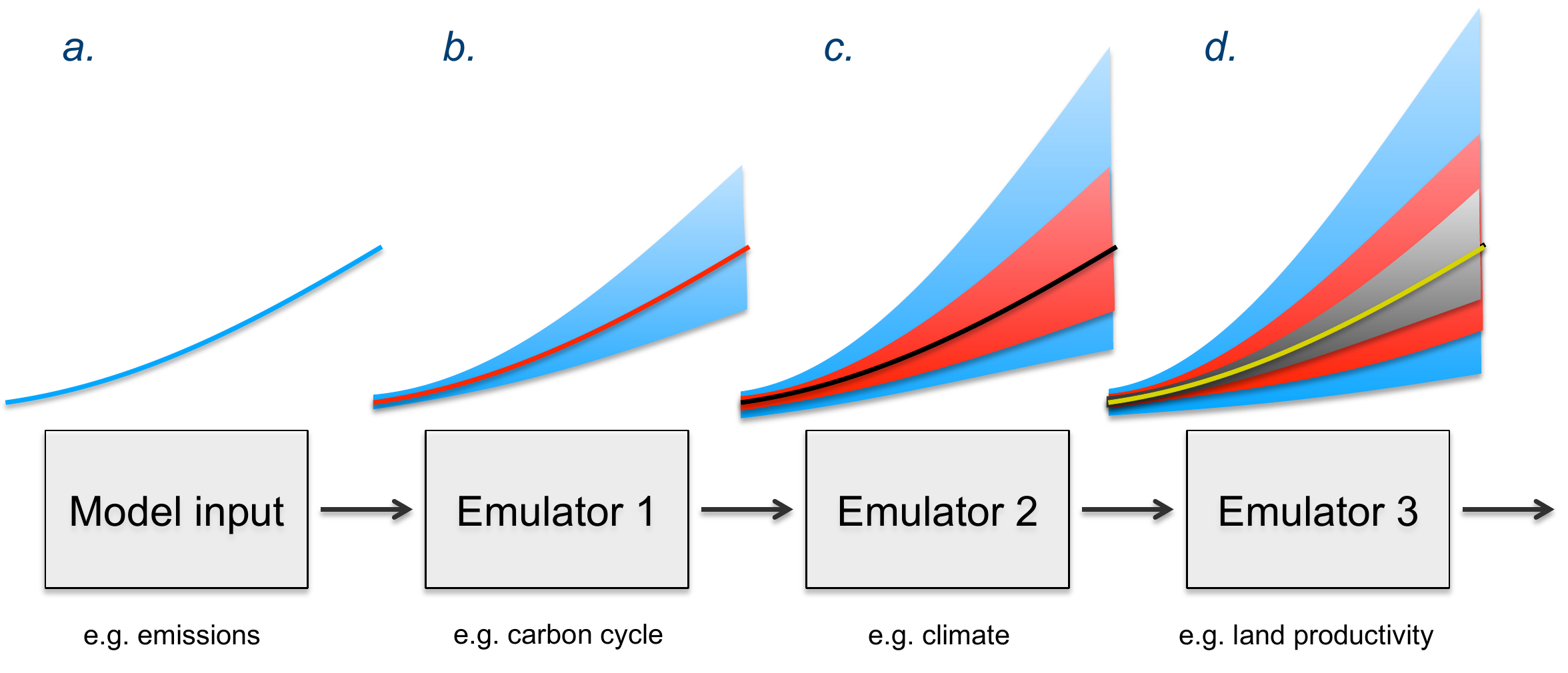}
	\end{center}
	\caption{Schematic representation of cascading uncertainty across statistical model emulators.}
	\label{fig:Figure1}
\end{figure}

\subsection{Uncertainty analysis in human-environment interactions and climate change\label{sect:2.3}} 

Models based on complexity theory can better describe the complex interactions between the socio-economic and the natural (environment) systems. Such interactions are at the heart of climate policy, and they can be effectively simulated at a low cost through the use of statistical emulators.

Feedbacks between the economy and the natural environment take place through direct human intervention (e.g. land-use change) or indirectly through the impact of economic activity (e.g. the generation of air pollution or the emission of greenhouse gases causing climate change). The natural environment influences, in turn, human action leading to a highly complex system of feedbacks. The allocation of land for agriculture, including crops for biofuels, interacts directly with the climatic system through phenomena such as deforestation, land-use change emissions and desertification, but also indirectly through the economy. The extensive use of water further constrains the availability of natural resources for human use, in what is increasingly referred to as the \emph{food-energy-water nexus}. At the same time, natural resource use stems directly from economic processes, involving the demand for agricultural, energy and forestry commodities that are traded in international markets.

Exploring complex interactions between the environment and the socio-economic system requires detailed representations of the role and behaviour of the natural environment, which is no minor endeavour. Indeed, climate modelling is carried out with the most powerful supercomputers currently available. The response of the environment to anthropogenic influence is increasingly well understood \citep{IPCCAR5WGI} and some of this knowledge has been reproduced reliably enough through the use of emulators (statistical representations). Emulators can be used directly, without requiring detailed parallel simulations with supercomputers alongside an economic model. Such statistical representations, which interpolate climate responses based on large climate model output data, have now been used for some time \citep[e.g.][]{Meinshausen2009}, facilitating the access by social and economic modellers to quantitative results from climate science, usually carried out in different institutions.

In this context, a frequently used tool is the linear approach of `pattern scaling' \citep{Tebaldi2014}. This approach assumes that the spatial pattern of change in the climate output of interest is invariant with respect to time and forcing. This approximation is often inadequate, however, for instance in the case of land-use change, which not only has significant impacts on the global climate through greenhouse gas emissions but also has a more localised (i.e. `pattern changing') impact on climate through changes in surface albedo, moisture transfer and river runoff (Myhre et al., 2013). A more general approach that allows for changes in the pattern of climate impacts has been developed recently \citep{Holden2010} and applied to `emulate' complex models of a range of climatic systems. The technique has been used to produce emulators for the climate model PLASIM-ENTS \citep{Holden2014}, the carbon cycle model GENIE \citep{Holden2013}, and the land surface model LPJmL \citep{Oyebamiji2015}. These emulators have already been applied in a range of integrated assessments \citep[including but not limited to][]{IEAWEO2013, Labriet2013, Mercure2014}.

Such a methodology provides a useful platform for the integration of large amounts of environmental data into socio-economic and policy analyses. By concatenating such emulators in `chains' (see schema in Figure~\ref{fig:Figure1}), one can indeed obtain a representation of uncertainty cascading across sectors. This technique avoids possible biases that may emerge from the use of median trajectories when linking different models. And it enables the exploration of representations of all likely trajectories simultaneously, at low computational costs. 

The cascading of uncertainty that increases with simulation timespan stems from the general property of complex non-linear models, which produce scenarios that diverge from each other exponentially for arbitrarily small changes in starting parameters (i.e. the sometimes termed `butterfly effect' in chaos theory), typical of climate models. This type of uncertainty is also a property of an economic model with positive feedbacks, as well as of diffusion-based technology models. More generally, it would be inconsistent to treat economic projections as deterministic (e.g. a GDP trajectory in equilibrium), while treating climate projections as probabilistic (e.g. 95\% probability range for a global warming trajectory). It is, however, entirely possible to assign `uncertainty bounds' to projections of a socio-economic model through the use of statistical analysis. Thus, the uncertainty faced in climate policy-making can be effectively tackled through such statistical shortcuts. Depending on the sensitivity of each sub-system to perturbations (i.e. the rate of divergence of scenarios, and conversely, the possible presence of `attracting states') the cloud of uncertainty may increase moderately or dramatically as it propagates through the chain. This in turn provides a clear quantification of our ability to reliably model the ensemble of systems.

\subsection{Policy effectiveness, behaviour and implementation\label{sect:2.4}} 

Tools from marketing research, anthropology and behavioural economics can be very useful to improve our understanding of consumer and investor behaviour. They provide a powerful addition to economy-environment modelling, namely the ability to predict quantitatively the most likely aggregate response to policy instruments. Valuable insights can be gained in this way. In climate change policy, which requires action sooner rather than later, there is significant value in the ability to predict the effectiveness of emissions reduction measures. However, assessing such feasibility requires knowledge of their legal and political implications and constraints (e.g. political feasibility, legal consistency), which may be just as important. Apparently minor differences in the applicable legal framework (e.g. the modalities for the acquisition of land, the modalities for the initial distribution of emission reductions or for their banking, the local content requirements added to a feed-in-tariff scheme to make them politically palatable) may, in fact, make certain developments less effective, or more vulnerable to challenge, or even block them entirely. 

Much policy analysis has been carried out in connection with specific policies proposed to address externalities such as pollution or environmental degradation, including taxes and other financial incentives imposed on households, firms or consumers. The justification for taxes is often based on considerations of ethics and social justice \citep[][Ch. 3]{IPCCAR5WGIII}; it less frequently results from the analysis of their likely effectiveness \citep[][Ch. 15]{IPCCAR5WGIII}. Indeed, policies are often chosen and adopted without much prior knowledge of their likely effectiveness, which ultimately lies with investor and consumer decisions. Their impact is mostly assessed \emph{ex-post}; however here, relying on ex-post policy evaluation does not fit in the timescale of action.

Policy analysis for climate change mitigation is complex, and realistic proposals must take into account the political and institutional (legal -- both domestic and international) context. For example, some environmental differentiation techniques (e.g. certain subsidies or feed-in tariffs), may be inconsistent with international economic law and can attract significant legal difficulties \citep{Vinuales2012}. Furthermore, the introduction of carbon pricing mechanisms (e.g. a cap-and-trade system where the allowances are freely allocated -- at least initially -- on the basis of prior emissions) can be politically very different from that of energy/fuel standards or emissions taxation, and it would therefore have a different likelihood of success. Thus, studying the effectiveness and feasibility of policy measures requires integrating expertise on the mechanisms of policy-making and law with expertise on the modelling of decision-making by investors and consumers. However, such a broad integration of expertise is extremely rare in contemporary state-of-the-art climate change mitigation research.

We propose here, as part of the working structure of simulation-based sustainability policy assessments, a two-way feedback with knowledge of domestic/comparative politics and environmental, investment and trade law. This would effectively guide the `what if' approach to scenario-creation for testing potential policies, as opposed to proposing policies that are already `optimal'. Effectively, in a model that is not based on optimisation, policies are assessed on the basis of their ability to effectively achieve certain objectives through the simultaneous use of several policy instruments that interact with one another. Such an approach avoids the more common siloed assessments, and it accounts for policy impacts across sectors. This approach is in fact recommended by the European Commission in its Impact Assessment guidelines \citep{EC2009, EC2015}.

\section{Practical relevance of the paradigm shift: specific applications to four key climate policy issues\label{sect:3}}

\subsection{The impact of consumer heterogeneity on the diffusion of new transport technology \label{sect:3.1}}

\begin{figure}[t]
	\begin{center}
		\includegraphics[width=.8\columnwidth]{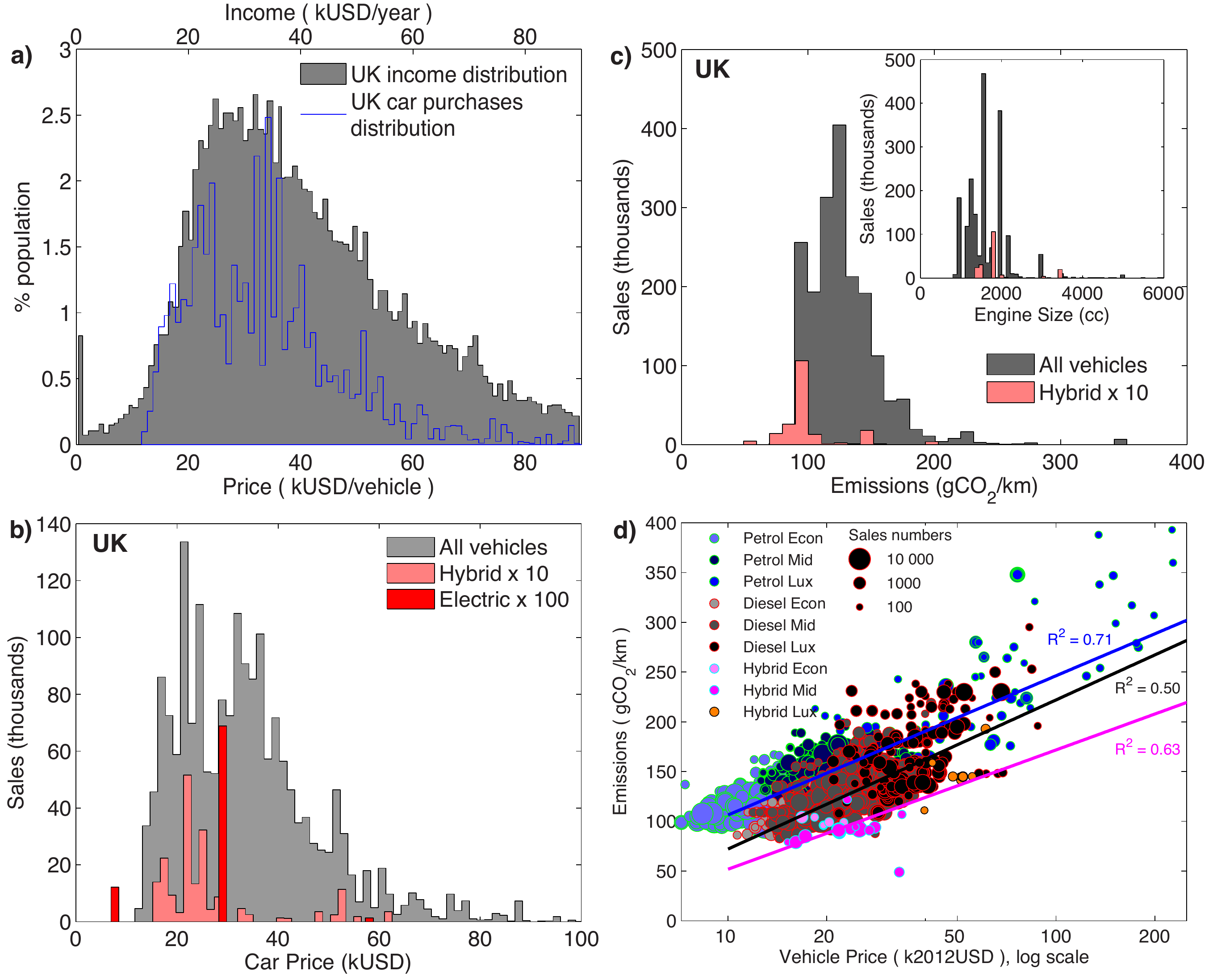}
	\end{center}
	\caption{Data relating the impacts of consumer diversity and market structure on the effectiveness of emissions taxation in the private personal mobility sector \citep[reproduced from][]{MercureLam2015}. a) Comparison of the UK income distribution to the price distribution of recent vehicle purchases. b) Comparison of UK market coverage between conventional and unconventional vehicle engine technology. c) Associated distribution of vehicle carbon intensities. d) Structure of the UK vehicle market for prices and emissions from which likely effectiveness of taxes can be deduced. These properties were found to vary across economies \citep{MercureLam2015}.}
	\label{fig:Figure2}
\end{figure}

The importance of agent heterogeneity for technology adoption can be empirically established, as shown in figure~\ref{fig:Figure2}, using private passenger vehicle purchases as an example derived from recent work \citep{MercureLam2015}. As shown in figure~\ref{fig:Figure2}, in the UK, the distribution of car purchase prices follows closely the income distribution (panel a). Cars of different prices have different manufacturer-rated emissions, which are thus similarly distributed (panel b). Vehicles with alternative engine technologies have yet a different distribution, and this stems from their gradual process of diffusion, which takes place unevenly with respect to the distribution of conventional vehicles (panel c). Finally, rated emissions are correlated with vehicle prices through a log-linear relationship. 

Car purchase choices largely depend upon each consumer's respective social group, through social interactions, as has been shown in empirical work \citep[for example][]{McShane2012}. This generally explains the relationship between consumption behaviour and income distribution shown above. Consumers do not attempt to minimise their transportation costs; instead they apparently purchase what is most common in their visible surroundings, and the diversity of social groups is what forms the lognormal distribution shown above. Alternatively, we can say that consumers maximise utility within a subset of the market (bounded rationality) defined by their social interactions. 

It thus appears necessary to look at each band of income, and each type of consumer, separately. If we now explore possible substitutions between models that could result from emissions reduction policies in the car market, we can calculate elasticities of substitution from these distributions by determining statistically which vehicle models are most likely to be chosen within the same new price band (after tax). Therefore, the rate of adoption of consumer technology, including low carbon systems, stems precisely from this diversity, which varies across the world \citep[as shown in][]{MercureLam2015}. This enables us to determine the effectiveness of certain low-carbon policies (taxes, subsidies) at incentivising technology substitution, using market data. In point of fact, this is exactly what marketing research does to forecast sales before placing new products in the market.

Such information can be fed into technology diffusion models that aim to reproduce typical S-shaped profiles \citep[e.g. as in the Future Technology Transformations (FTT) model,][]{Mercure2012} to provide expected rates of adoption (see Figure~\ref{fig:Figure4} further below). Such a representation reproduces technology lock-ins. This picture is incomplete, however, as it does not represent attitudes and culture. Yet, just as with firms attempting to place new products in the market, the more detailed information thus gained helps to better characterise the rates of technology adoption that could result from proposed policies throughout the diffusion cycle. By contrast, optimisation-based models would typically characterise variations in consumer behaviour, at best, by parameterising different discount rates for technologies attributed to particular market segments.

\subsection{Green growth: employment and income impacts of low-carbon investments\label{sect:3.2}}

\begin{figure}[t]
	\begin{center}
		\includegraphics[width=1\columnwidth]{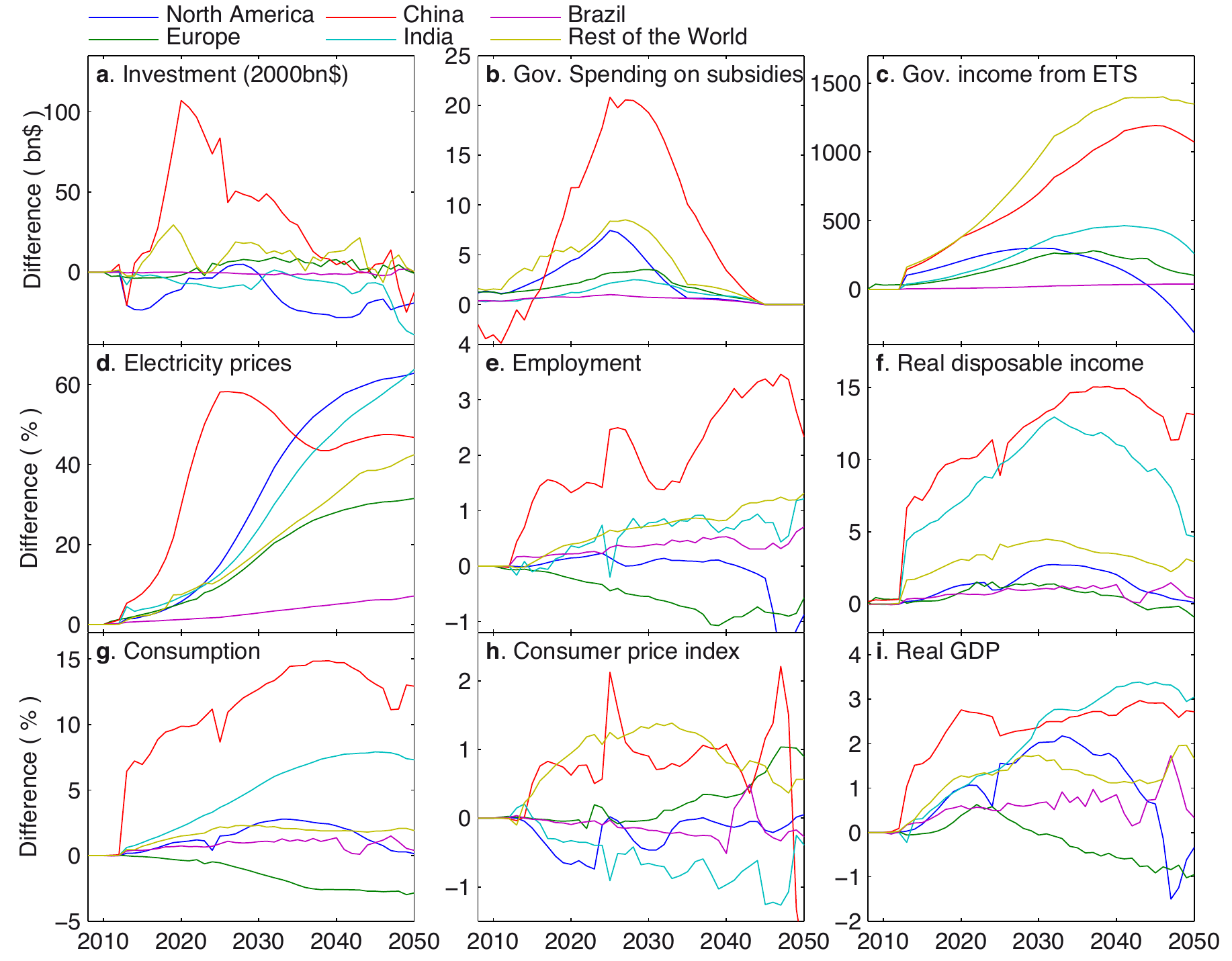}
	\end{center}
	\caption{Example of possible economic impacts of investment in low-carbon electricity generators, when studied using a non-equilibrium post-Keynesian model, E3MG-FTT \citep[see][]{Mercure2015b}. Vertical axes refer to changes from the baseline, in billion current USD for the top three panels, and in percentage points for the six lower panels.}
	\label{fig:Figure3}
\end{figure}

The importance of accounting for the interactions between lenders' expectations, technology diffusion and macroeconomic variables can be illustrated by reference to the link between low-carbon investments, income and employment. Figure~\ref{fig:Figure3} is based on a non-equilibrium macroeconomic model, which is not subject to the full employment constraint \citep[E3ME/E3MG][]{E3ME}. It shows a possible causal chain in the process of technology finance based on expectations impacting the economy in the electricity sector. This calculation is further described in \cite{Mercure2015b}.

In this model, no limit is imposed on finance for technology developments, which in contrast to an equilibrium model, has no direct relationship to interest rates (assumed in equilibrium to clear the money market). In other words, financial resources are given to entrepreneurs by banks for low-carbon investments through credit creation. The model's critical assumption is only the solvency of entrepreneurial activity, i.e. economic viability of low-carbon projects. Therefore it is not claimed that the economic system has no limit at all on money flows: it is assumed that all technology ventures that are financed in any scenario are profitable (for instance by assuming sustained credible policy and/or prices that make these ventures feasible), and that government and/or private debt is not indefinitely increased. This ensures that situations that could lead to financial instability through unsustainable debt growth are not created. This example is selected to show that economic outcomes of mitigation policy in models are entirely dependent on model architecture, and that impacts can be (but are not necessarily) beneficial if one allows for non-equilibrium effects.

In this example, the electricity sector is decarbonised by 90\% \citep[scenario previously published in][]{Mercure2014}. Higher costs of low-carbon electricity generation technologies (e.g. wind turbines, solar panels, carbon capture and storage) are passed on by utilities into their bills to customers, i.e. into higher prices of electricity, proportionally to the evolving technology composition. Entrepreneurial activity requires loans to finance new low-carbon capital, which requires more investment than existing fossil fuel generators. \emph{Banks create money to finance these ventures}. Loans are paid back over capital lifetime by firms using a higher price of electricity (and, for example, possible feed-in tariffs), and \emph{the additional money is gradually destroyed} as the loans are paid back. In the model, as shown in figure~\ref{fig:Figure3}, the higher level of money flows from investment (panel a) creates jobs (panel e), increases disposable income (panel f) and consumption (panel g), with possible effects on inflation (panel h). Meanwhile, higher prices of electricity increase operational costs of most sectors, and thus decrease employment, household income and consumption (same panels), i.e. an offsetting force. These two effects were observed to roughly cancel each other in the model and scenario \citep[see][]{Mercure2015b}. A positive `green growth effect', in this particular case, is generated in parts due to redistribution policy, in parts due to increased employment. Revenue raised by fuel taxes (panel c) aimed at incentivising technological change (carbon pricing), minus spending on technology subsidies (panel b), is indeed recycled to lower income tax. This moves the system's balance towards a higher income level than in a baseline scenario (higher GDP, panel i). This effect subsides in later years (US, Europe) when investment and redistribution declines, while the price of electricity remains high, and the effect may even reverse when the technology transformation is completed and society faces servicing debt only. Debt servicing takes place through consumers paying a higher price for electricity.

Note, therefore, that while economic growth is enhanced in this scenario, private debt is also increased beyond the end of the simulation. In non-equilibrium theory, borrowed investment flows indirectly contribute to increased aggregate demand in the short run, when loans are issued, and to decreased demand in the long run, when loans are gradually paid back. Increasing the level of borrowing generates debt-based growth, which if done indefinitely, generates significant prosperity but eventually leads to collapse through a financial crisis. It is unlikely that climate change mitigation would lead to indefinite borrowing. But it will likely require significant amounts of finance.

According to Keen \citep[e.g.][]{Keen2011}, extended debt-based growth and excessive debt levels have been the underlying cause of the recent banking crisis, and possibly many other economic cycles historically \citep{Perez2001}. This points to a complex entanglement between strategies for economic recovery after the economic crisis, and strategies for climate change mitigation policy. The financial crisis involved banks refusing to lend despite quantitative easing, while mitigation, a potential economic stimulant, requires increased amounts of finance in the energy sector.

We conclude that further research is required to understand the levels of debt and risk, private and public, that economies can realistically undertake to reduce global emissions, as well as to clarify the collective expectations on their economic returns. In a complexity/non-equilibrium perspective, the impact of collective expectations appears to be the key issue to explain, rather than simple welfare effects of different allocations of fixed amounts of capital typically analysed with current optimisation models. In the context of scenarios of rapid decarbonisation, the analysis of debt and access to finance becomes particularly crucial.

\subsection{Model integration across processes with cascading uncertainty\label{sect:3.3}}

\begin{figure}[t]
	\begin{center}
		\includegraphics[width=1\columnwidth]{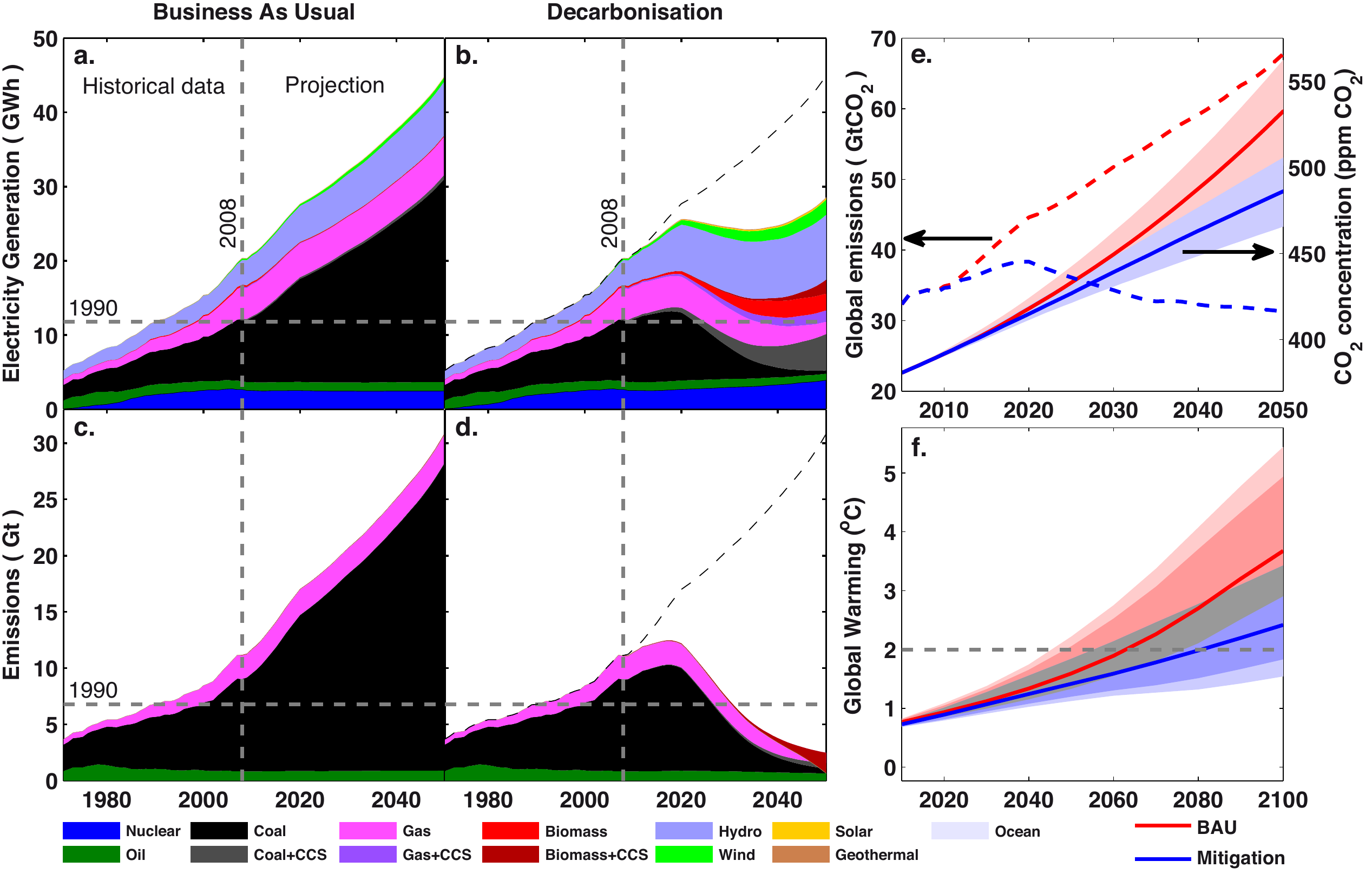}
	\end{center}
	\caption{Example calculation of the environmental impacts of electricity policy instruments using E3MG/E-FTT (left 4 panels) with OU emulators (right panels), from policy to global warming, with cascading uncertainty bounds from combined carbon cycle and climate system emulators.}
	\label{fig:Figure4}
\end{figure}

Large models of the natural world have many imperfectly known parameters as well as -- through multiple combinations of those parameters -- a considerable number of possible output values arising from variations in those parameters (even 10 settings of only 10 parameters would generate 10$^{10}$ possibilities). Statistical modelling techniques are required to interpret such a large space of uncertain outcomes. When several models are `soft-linked' in a causal chain, the uncertainty of models upstream inevitably generates higher uncertainty downstream as we give ever wider ranges of parameters to models. Figure~\ref{fig:Figure4} gives an example of this with 3 soft-linked models, starting with individual emissions scenarios from E3MG-FTT \citep[note that these scenarios are the same as in Figure~\ref{fig:Figure3}, their emissions trajectories coupled to climate emulators; they are also the same as scenarios a. and j. in][]{Mercure2014}.

In (a.), we have a baseline scenario of the composition of the global electricity sector, calculated under 21 regions independently. Using a composite scenario of emissions reduction policies that include carbon pricing, technology support policies (subsidies and feed-in tariffs) and regulations \citep[available in][]{Mercure2014}, the electricity sector is transformed towards low-carbon technologies (b.). Baseline emissions are projected to increase by 318\% based on their 1990 level (c.), while in the decarbonisation scenario, they are reduced by 90\% (d.).

These emissions scenarios are fed to the carbon cycle emulator GENIEem, which generates GHG concentration outputs with 95\% probability ranges (e.). These are then fed to the emulator of the climate model PLASIM-ENTS, producing a set of scenarios for global warming (f.), as well as other locally resolved changes of climate (on a 0.5o grid, not shown). The soft-linking thus produces double uncertainty ranges from the concatenation of uncertainty. We find that decarbonising the electricity sector by 90\% is not sufficient to avoid exceeding 50\% chance of exceeding the international target of 2$^{\circ}$C \citep{Mercure2014}. All sectors of the energy end-use system must be involved, notably transport. Cascading uncertainty ranges in this context is as important as cascading median trajectories.

\subsection{Cross-sectoral impacts of biofuels policy\label{sect:3.4}}

In an optimisation model of land allocation, best allocations of farming activities are determined across the surface of the land studied, which extends to the whole world for global models. This implies that farmers know instantaneously what is too little and too much, and never generate excess product. Without excess product or shortages, price fluctuations cannot occur. This is, of course, not what is observed \citep{Piesse2009}. However, building a model that can reproduce observed price fluctuations is a significant challenge, even though such fluctuations seem crucial to understand environmental degradation and biodiversity loss. Indeed, some deforestation occurs not directly because of an increasing consumption of agricultural commodities globally, but as a result of commodity prices fluctuating globally, due to self-reinforcing expectations of returns \citep[e.g.][]{Arima2011, Morton2006}. 

In a non-equilibrium model that includes representations of decision-making by heterogeneous agents in agriculture with expectations of return on their crops, in tandem with a global model of the economy and bilateral trade of agricultural commodities, very different dynamics can emerge. The use of heterogeneous agent functional types has been suggested to improve model realism \citep{Arneth2014, Rounsevell2014}. While price substitutions and changes in trade patterns can occur with the consumption of agricultural commodities, massive land clearance and conversions can also take place to accommodate expectations around changing prices. Such models generate highly complex dynamics, but are useful to determine what types of transformations could occur in the future without appropriate land management policy. Complex dynamics however will only arise in models that incorporate heterogeneity, increasing returns and expectations. When land-use decisions are based on expectations and influenced by neighbourhood effects, then similar contagion dynamics as in technology diffusion may take place, i.e. the diffusion of agricultural practices.

When brought into the picture, biofuels policy adds further uncertainty and complexity. On the one hand, the willingness to pay for ethanol in some parts of the world could outbid the ability to pay for food commodities in other parts of the world, while, on the other hand, shortages of food may also be felt through commodity prices after land-use decisions are already made and applied. Effectively, some biofuels policies have the potential to open the door for substantial fluctuations in food prices, but we do not know which ones, and at what scale. This issue requires models of a type that is currently not available, but needs to be addressed rapidly.

\subsection{Is higher model complexity a bonus or a drawback for policy-makers?\label{sect:3.5}}

Lower complexity or reduced-form models are often argued to be advantageous to use due to their higher apparent transparency, and thus more appropriate for policy-making. An example is the use of Nordhaus' DICE model, which was originally designed for illustrative purposes using sketchy data, but has been used for policy-making purposes in some countries, including the USA and the UK. 

Lower complexity \emph{does not mean better science}, and simplifications can, in some cases \citep[e.g. as with][]{Nordhaus2010} lead to potentially plainly incorrect conclusions (its single sector/good representation of the economy excludes many important effects such as spill-overs and multipliers). Moreover, and more fundamentally, the use of deterministic model projections (e.g. policy-making based on an optimal carbon price that results from intersecting deterministic curves for the social cost of carbon and the marginal abatement cost) is not a scientifically correct methodology. It is clearly inconsistent with the way complexity-based climate modelling is conducted. No one would realistically claim the ability to predict the exact mean global warming in 2050, and there is no reason why such claims would be justified in economics. 

We thus argue that while higher complexity models may be more difficult to use in policy circles, such difficulty is clearly offset by their greater realism and rigour. The key consideration is not whether simpler or more complex models should be used, but whether the \emph{science-policy interface} is capable of relaying the results of more powerful and realistic models to policy circles.

\section{Conclusion: a world of new possibilities for sustainability policy-making\label{sect:4}}

Equilibrium and optimisation-based models are appropriate to use for normative exploration and identification of desirable future configurations of the technology-economy-environment system. Given the fact that they are comparatively highly detailed and tested, they are currently treated as the standard approach. This is possible because normative analysis relies entirely on assumptions and does not require empirical knowledge of actual human behaviour. However, such models support only certain steps of the policy cycle, as they provide ambiguous information regarding how to achieve -- through policy interventions -- the ideal configurations they portray or how likely they may be. This gap is a direct consequence of a \emph{lack of causal relationships with human behaviour}. Producing scenarios that accurately forecast the future course of events as a result of policy choices, with any likelihood (however low), requires fine-grained representations of human behaviour, its diversity and multi-agent interactions. These representations are as necessarily imperfect as they are necessary to come close to real life.

It is often argued that forecasting is not possible, and that serious sustainability science can only express itself in the form of exploratory scenarios \citep[e.g.][]{vanVuuren2011}. We submit in contrast that forecasting is both necessary and inevitable, and can be improved within the bounds of finite predictability with increased attention to known nonlinearities and interaction effects \citep{Tetlock2015}. The climate sciences provide a major example of an area where outcomes are expressed in terms of likelihood levels based on model statistics. Forecasts are not always accurate, but they are nevertheless useful due to their careful quantification of probabilities, parameters and their uncertainty. There is no inherent reason why this is not possible in the social sciences. We note that the climate community now prefers the term 'projection' to indicate that model simulations are predictions that are conditional on their inputs. This distinction of terminology, while relevant to the remaining exogenous assumptions in our case, does not affect our fundamental conclusion.

The main challenge lies in the quantification of likelihood, which is not possible with current mainstream socio-economic models. We do not claim, however, to be able to predict the onset of wars, election results, natural disasters, strategic political choices or other unique events. Thus, an important caveat must be made, in that our methodological approach excludes the occurrence of some events (just like in weather forecasts, which do not take into account possibilities of volcanic eruptions, even in active areas). 

Yet, the modelling paradigm shift proposed here opens a large spectrum of possibilities for sustainability policy-making. It allows for the quantification -- within uncertainty bounds -- of the expected effectiveness of specific policies aimed at inducing particular agent groups to take particular decisions (e.g. consumer purchases, investment choices, land-use decisions). The ability to conduct such forecasts entails significant advantages, including the reduction of the uncertainty involved in creating policy portfolios, and the possibility to explore their impacts across sectors through the coupling of several sectoral models. For example, one could study the impact of technology support policy for electric vehicles on electricity prices, which would depend on the pace of their adoption, or the impact on food prices of large scale land regulations to protect biodiversity or, still, the impact on deforestation of biofuels policy in transport. These questions present major analytical challenges, but their understanding cannot wait any longer. We believe that the new generation of non-equilibrium models proposed in this article is capable of rising to this challenge.

\section*{Acknowledgements}

The authors wish to thank the full GUIDEPOST team for highly fruitful discussions over the length of funding proposal developments. The authors also warmly thank assiduous referees for their comments that led us to greatly improve our manuscript, and A. Jarvis, whose comments made us revisit concepts in detail. J.-F. M. thanks D. Crawford-Brown, T. Barker and 4CMR staff and students for support. J.-F. M. and H. P. thank Cambridge Econometrics Ltd staff for support. J.-F. M. acknowledge the UK Engineering and Physical Sciences Research Council (EPSRC), fellowship no EP/ K007254/1 and J.-F.M. and J. V. acknowledge a networking grant of the EPSRC (Newton Fund) EP/N002504/1.

\section*{References}

\bibliographystyle{elsarticle-harv}
\bibliography{../../CamRefs}

\end{document}